\documentclass
[letterpaper]{jpconf}
\usepackage{
graphicx}

\begin{document}

\title{
{\footnotesize\rm\hfill IIT-CAPP-11-10}\\ 
Study of multiple scattering in high magnetic fields}
\author{Daniel M. Kaplan$^1$ and Thomas J. Roberts$^2$}
\address{$^1$ Illinois Institute of Technology, Chicago, IL 60616, USA}
\address{$^2$ Muons, Inc., Batavia, IL 60510, USA}

\begin{abstract}
Muon cooling for a neutrino factory or muon collider can be achieved using low-$Z$ absorbers in strong focusing fields. Proposed cooling lattices place absorbers in solenoidal fields ranging up to 30 to 40\,T. The cooling performance of these lattices is determined by the interplay of ionization energy loss and Moli\`ere scattering, but Bethe's classic treatment of Moliere scattering ignores the helical motion of charged particles in solenoidal fields. When this motion is taken into account, the performance of these lattices can be better than predicted by simulations using the standard treatment.\\[.3cm]
{\sl Contribution to 
NUFACT 11, XIIIth International Workshop on Neutrino Factories, Super beams and Beta beams, 1-6 August 2011, CERN and University of Geneva\\
(Submitted to IOP conference series)}

\end{abstract}

\section{Introduction}
The question of the effect of strong magnetic fields on multiple scattering was raised by Lebrun in 1999~\cite{Lebrun}. He realized that Bethe's derivation~\cite{Bethe} of the Moli\`ere scattering distribution implicitly assumed straight-line motion between scatters. In contrast, in a strong magnetic field, a charged particle executes circular or helical motion, and in the limit of infinitely strong field, such a  particle's path is confined to a magnetic-field line. Clearly, this must reduce the multiple-scattering-induced spread in position as a charged particle traverses a scattering medium, as Lebrun demonstrated using Geant 3. Since this is exactly the configuration proposed in many implementations of muon cooling, simulations of muon-cooling channels risk underestimating their effectiveness, especially as the beam emittance shrinks and the field strength grows to 20\,T and beyond~\cite{final-cooling}.

In order to understand the degree of underestimation that may occur in such simulations, we undertook to compare the treatment of scattering in magnetic fields in the  principal codes used in recent MAP and NFMCC cooling studies: G4beamline~\cite{G4BL} and ICOOL~\cite{iCool}. To provide a simple and well-understood baseline, we also evaluated the effect using a Gaussian approximation to the scattering distribution.\footnote{For simplicity, we disabled $dE/dx$ energy loss in all three calculations.}

\section{Gaussian Model}

\begin{figure}
\hspace{.45in}\rotatebox{90}{\qquad\qquad\qquad\qquad~~~~\textsf{\footnotesize \# events}}
\centerline{\hspace{-1.2in}\includegraphics[width=2.25in,trim=35 420 335 135 mm,clip]{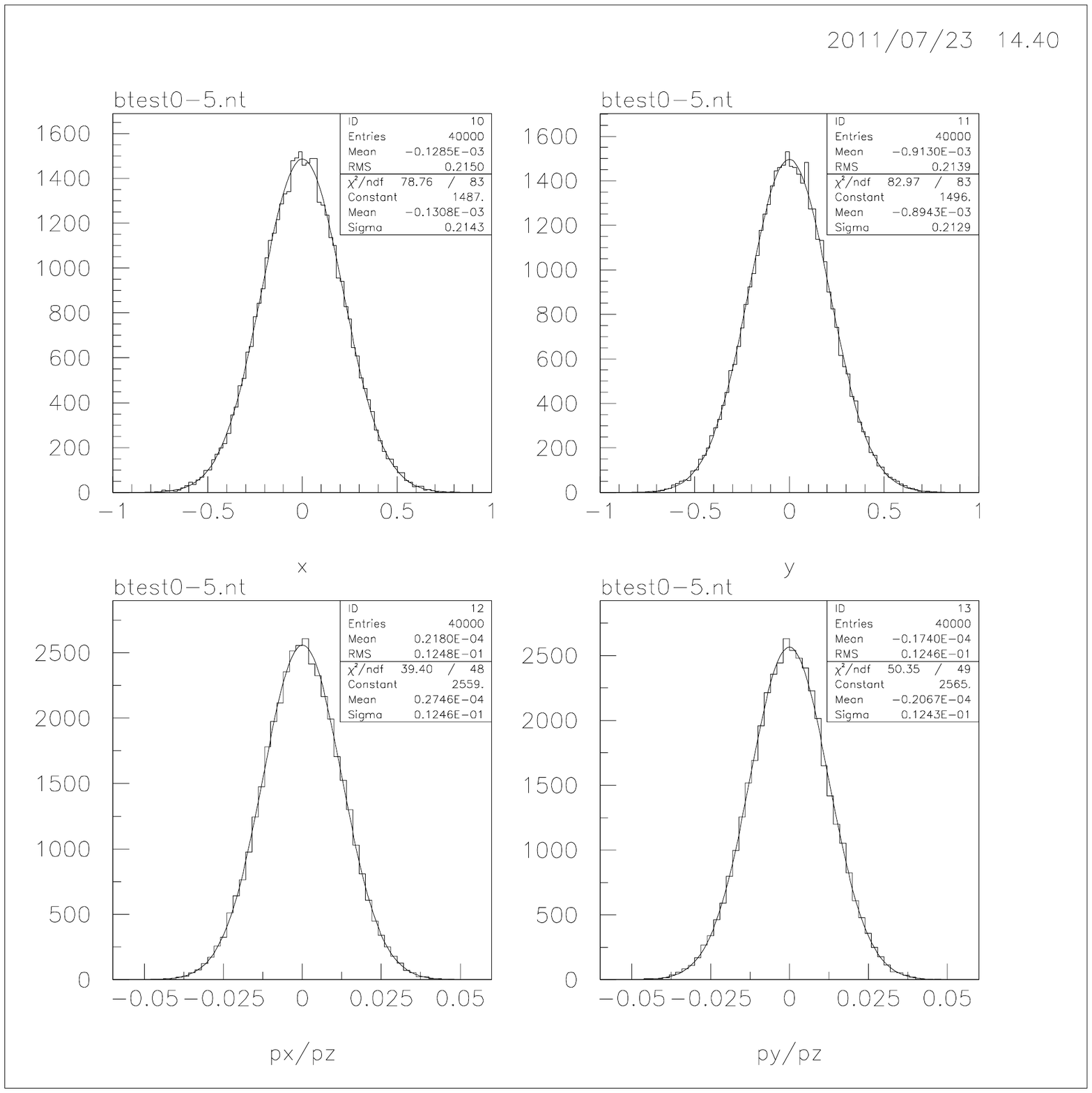}\,\includegraphics[width=2.25in,trim=35 170 335 385 mm,clip]{btest0-5}}
\centerline{
\includegraphics[width=2.25in,trim=35 435 335 135 mm,clip]{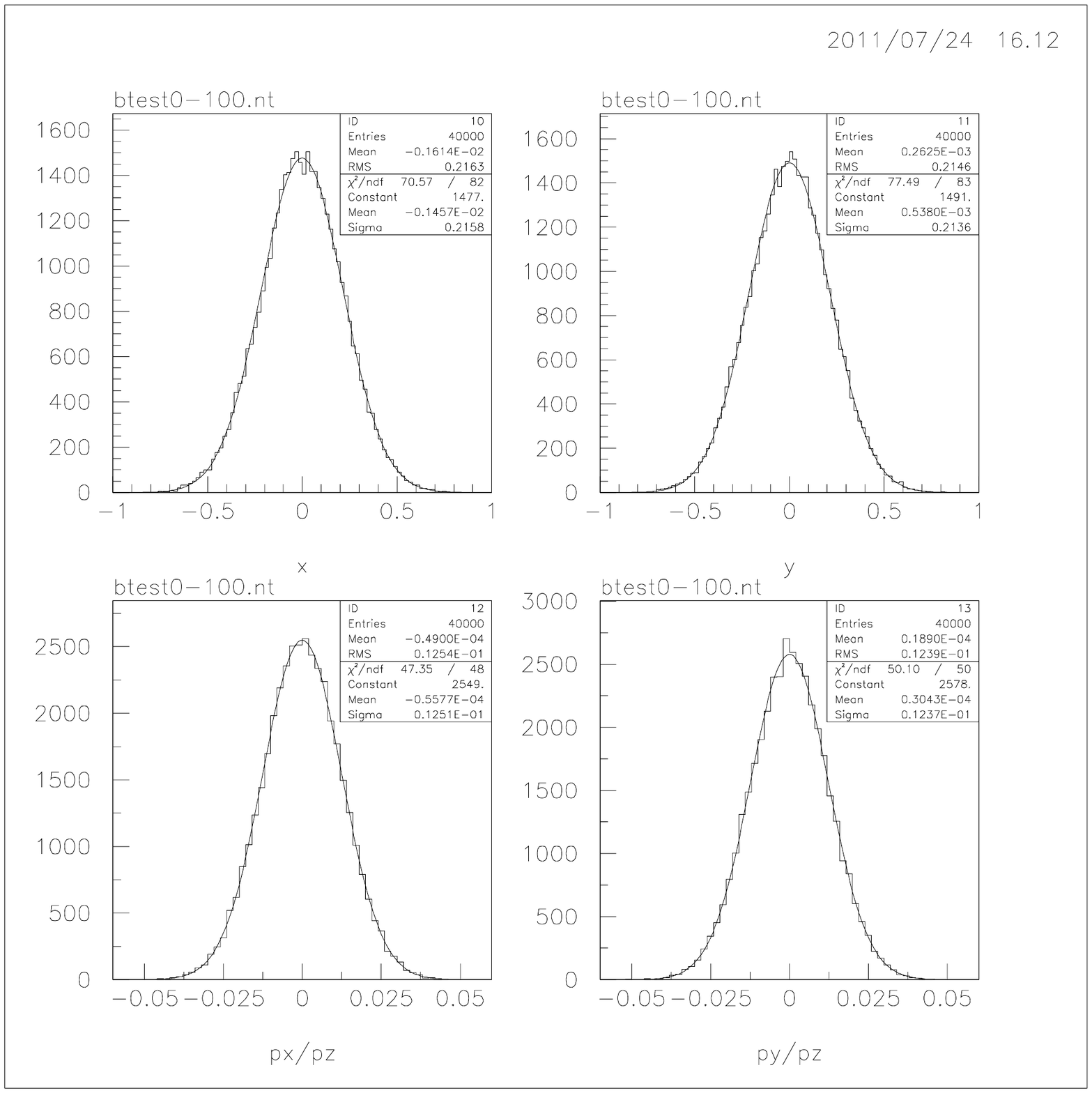}\,\includegraphics[width=2.25in,trim=35 185 335 385 mm,clip]{btest0-100}}
\qquad\qquad\qquad\qquad\qquad\qquad\qquad~~~\,\textsf{\footnotesize$x$ (cm)}\qquad\qquad\qquad\qquad\qquad~~~~~\textsf{\footnotesize$p_x/p_z$}
\vspace{-.025in}
\caption{Distributions in (left) $x$ and (right) $\tan{\theta_x}$ at the end of a 30\,cm LH$_2$ absorber in Gaussian multiple-scattering model, for (top) 5 and (bottom) 100 absorber slices.}\label{fig:btest0-5}
\end{figure}

In our Gaussian study, we let each muon start at one end ($z=0$) of a 30-cm-long liquid-hydrogen scatterer (a typical muon-cooling-channel absorber length), with longitudinal momentum $p_z=200\,$MeV/$c$ and transverse momenta $p_x=p_y=0$. We subdivided the absorber into thinner and thinner longitudinal slices, tracing each muon's path as a straight line (if $B=0$) or helix (if $B\ne0$) between the centers of successive slices.  At the center of each slice, we allowed $\theta_x$ and $\theta_y$ to change by random amounts, with r.m.s. variation~\cite{PDG}
\begin{equation}
\sigma_{\theta_x}=\sigma_{\theta_y}=\frac{13.6\,{\rm MeV}}{\beta cp}\sqrt{\frac{\Delta z}{X_0}}
\left[1+0.038\ln\left({\frac{30\,{\rm cm}}{X_0}}\right)\right]\,,\label{eq:scat}
\end{equation}
where $\Delta z$ is the width of a slice and $X_0 = 866$\,cm is the radiation length of liquid hydrogen~\cite{PDG}.
Figure~\ref{fig:btest0-5} shows an example of the transverse spatial and angular distributions at the end of the absorber for the cases $B=0$ and $n_{\rm slices}=5$ or 100. As expected, the number of slices makes no difference if $B=0$.

\begin{figure}
\centerline{\includegraphics[width=3.75in,trim=5 15 5 10 mm,clip]{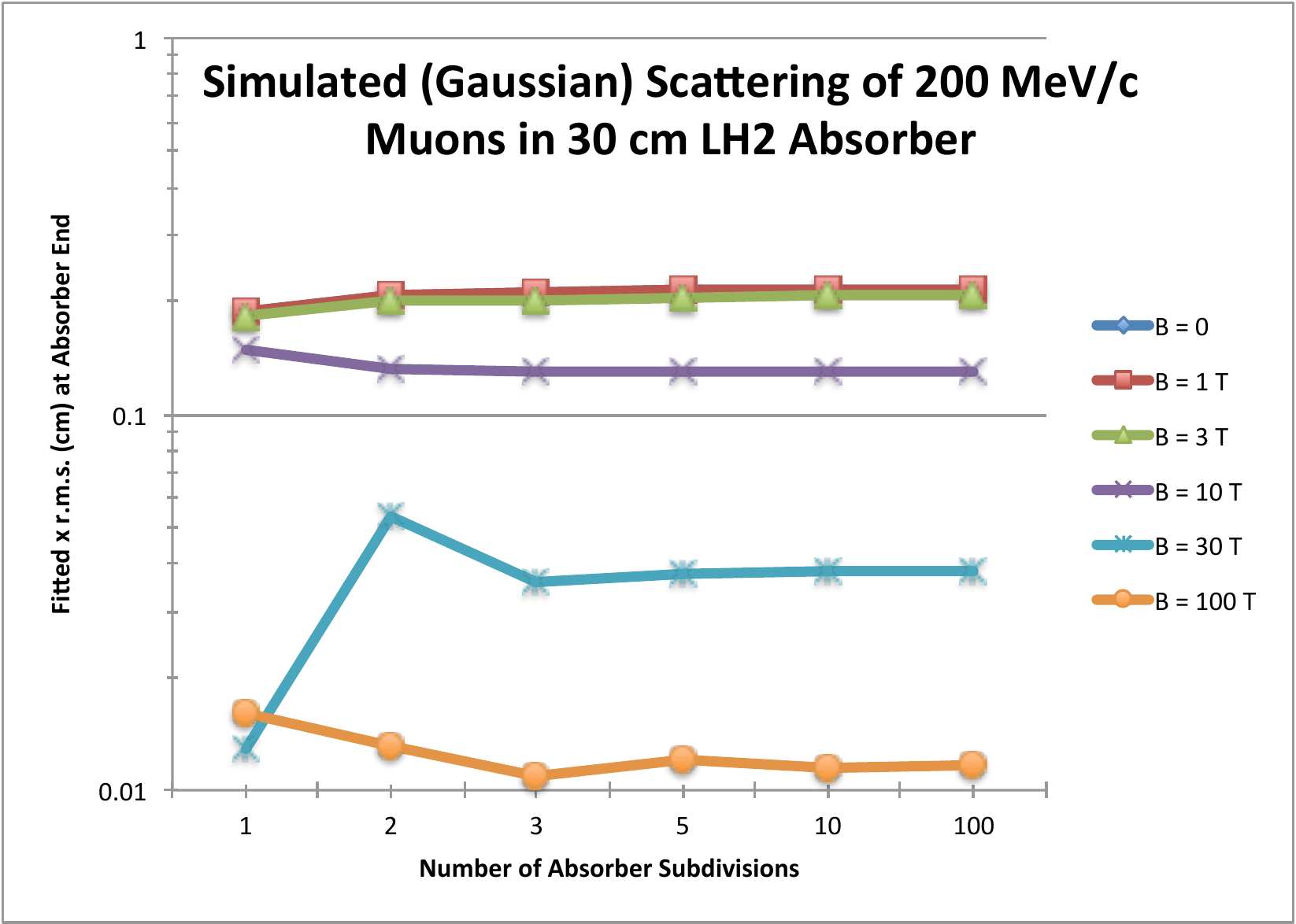}}
\vspace{-0.05in}
\caption{Root-mean-square value of $x$ after 30\,cm LH$_2$ absorber vs number of slices, in Gaussian model, for longitudinal $B$ fields indicated.}\label{fig:slices}
\end{figure}

We then allowed the longitudinal magnetic field to increase up to 100\,T and studied the number of slices required for the r.m.s.\ $x$ value at the end of the absorber to stabilize, finding that 10 slices suffice (Fig.~\ref{fig:slices}). We observe a dramatic suppression of spatial (but not angular) spread as the magnetic field increases, by a factor of $\approx$\,5 at 30\,T and $\approx$\,20 at (the admittedly impractical) 100\,T.

\section{G4beamline}

\begin{figure}
\rotatebox{90}{\qquad\qquad\qquad~~\,\textsf{\footnotesize \# events}}\hspace{-.15in}
\centerline{\includegraphics[width=2.9in,trim=0 24 2 0 mm,clip]{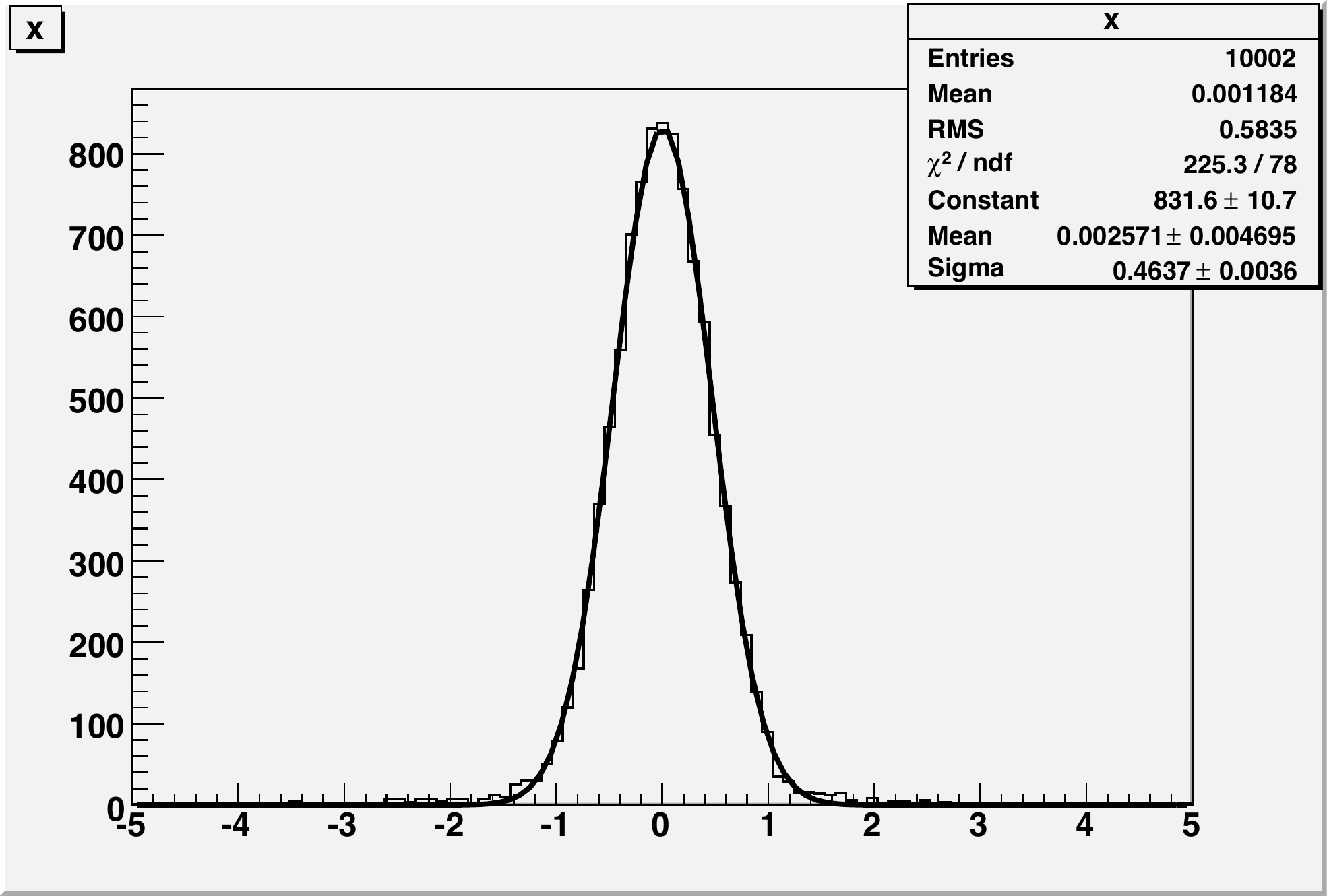}\includegraphics[width=2.85in,trim=0 23 10  0 mm,clip]{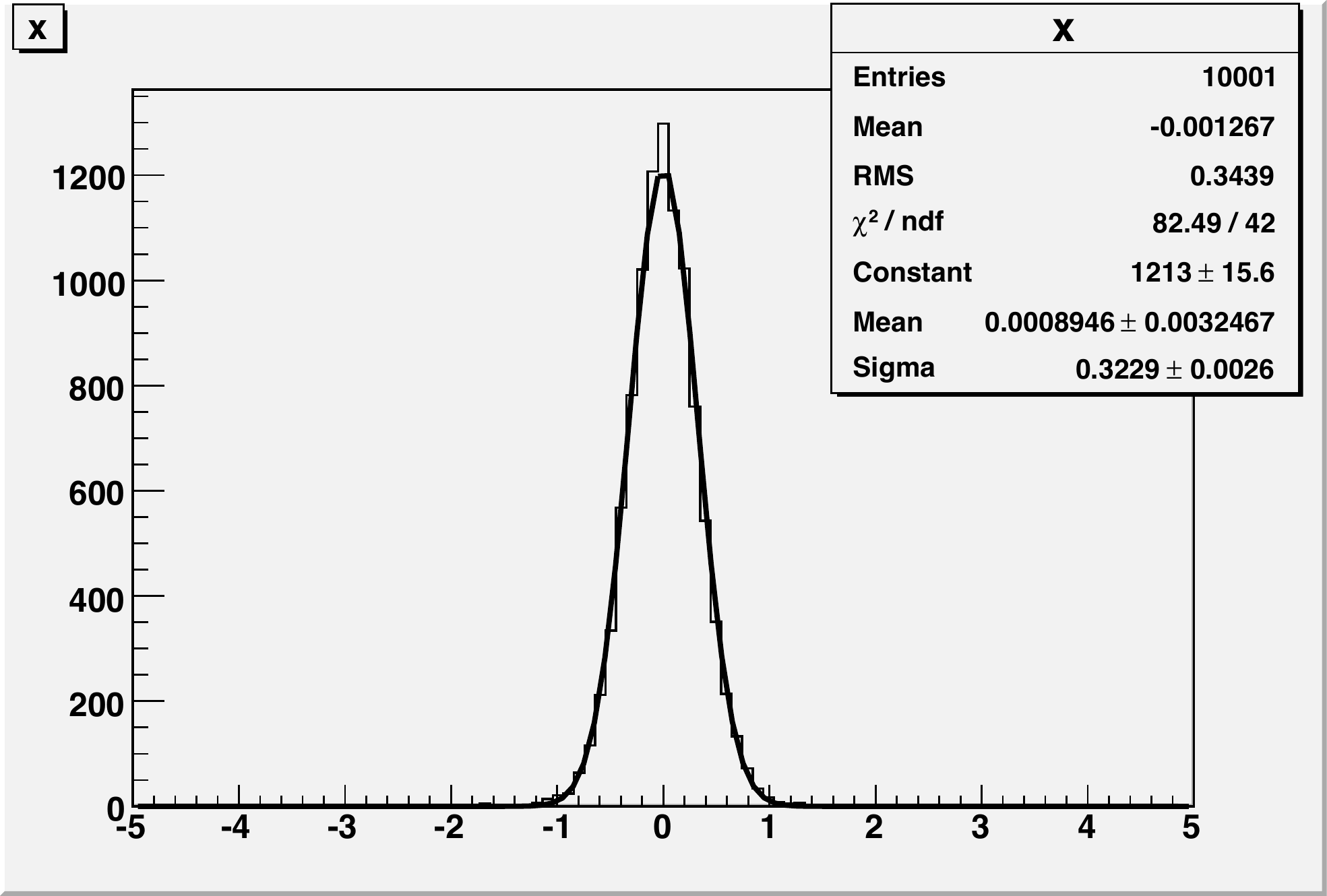}}\vspace{-.035in}
\qquad\qquad\qquad\qquad\qquad\qquad~~~~~\textsf{\footnotesize$x$ (mm)}\qquad\qquad\qquad\qquad\qquad\qquad~~~~~~~~\textsf{\footnotesize$x$ (mm)}
\vspace{-0.05in}
\caption{Distributions in $x$  after 30\,cm LH$_2$ absorber with $B=30$\,T, in G4beamline simulations, with (left) the default value maxStep\,$=100$\,mm and (right) maxStep\,$=1$\,mm.}\label{fig:G4BL}
\end{figure}

We next performed a similar study with the G4beamline simulation package. The ``knob" comparable to absorber subdivision is the choice of the maxStep parameter, which is 100\,mm by default. It represents the largest step along the particle trajectory that Geant 4 is permitted to take. Figure~\ref{fig:G4BL}  compares the r.m.s.\ $x$ distributions in a 30\,T longitudinal field after traversal of a 30\,cm LH$_2$ absorber by 200\,MeV/$c$ muons with  maxStep\,$=100$ or 1\,mm. The default value of maxStep leads to an overestimate of the r.m.s.\ spread by about 50\%.

\section{ICOOL}
\begin{figure}
\rotatebox{90}{\qquad\qquad\qquad~~\,\textsf{\footnotesize \# events}}\hspace{-.15in}
\centerline{\includegraphics[width=2.85in,trim=0 23 10 0 mm,clip]{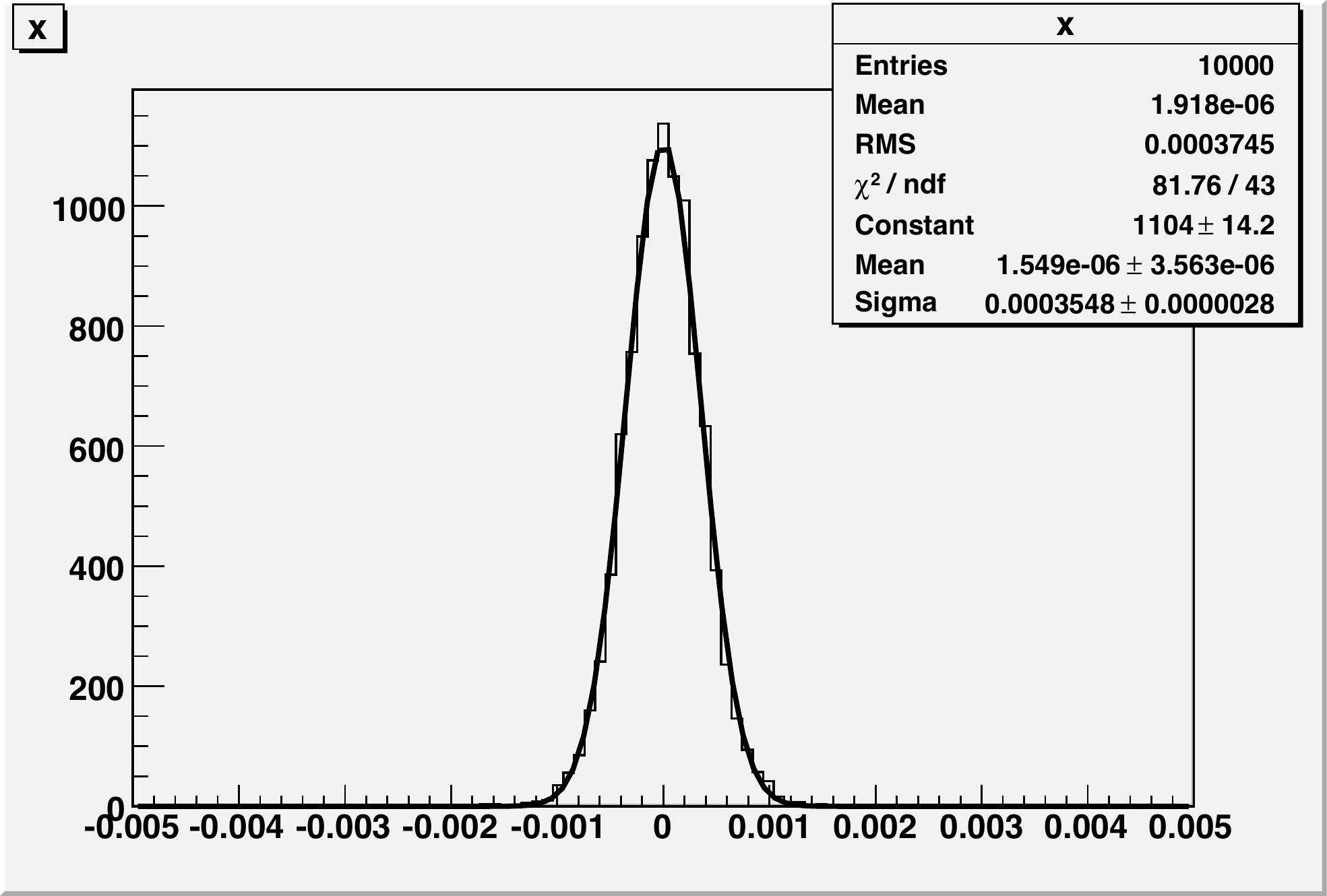}\includegraphics[width=2.85in,trim=0 23 10 0 mm,clip]{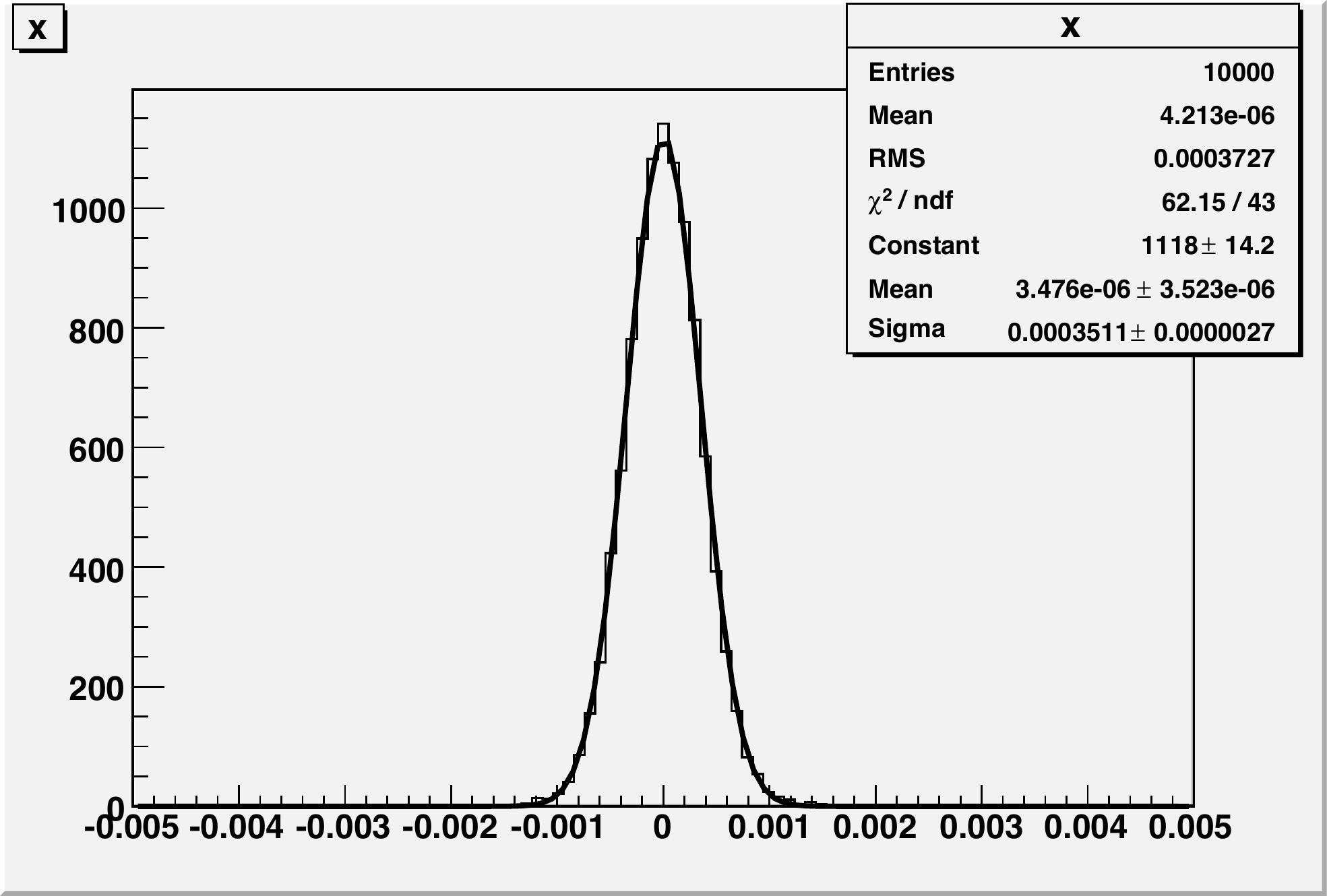}}\vspace{-.035in}
\qquad\qquad\qquad\qquad\qquad\qquad~~~~~~~\,\textsf{\footnotesize$x$ (m)}\qquad\qquad\qquad\qquad\qquad\qquad~~~~~~~~~\textsf{\footnotesize$x$ (m)}
\vspace{-0.05in}
\caption{Distributions in $x$  after 30\,cm LH$_2$ absorber with $B=30$\,T, in ICOOL simulations, with (left) ZSTEP\,$=10^{-1}$\,m and (right) ZSTEP\,$=10^{-3}$\,m.}\label{fig:ICOOL}
\end{figure}

Finally we studied the same situation using ICOOL. The relevant knob here is the choice of the ZSTEP parameter, which we took as $10^{-1}$ or $10^{-3}$\,m. In Fig.~\ref{fig:ICOOL} we see that the spatial spread is independent of the value of ZSTEP.

\section{Discussion and Conclusions}
We have verified Lebrun's conclusion that high magnetic fields suppress some of the effects of multiple scattering and can thus produce smaller beam emittances in ionization cooling than one might naively expect. Because of this effect, G4beamline cooling results at high fields are not reliable unless the maxStep parameter is chosen carefully. We have discussed this with Yonehara, who used G4beamline to simulate helical cooling channels at fields up to $\approx$\,15\,T~\cite{Yonehara}. We found that he performed a maxStep sensitivity study and so was not misled by this effect.
In contrast, ICOOL uses a variable-stepping algorithm by default, and so is
largely insensitive to the user's choice of ZSTEP. Thus Palmer's simulations of 30--40\,T ``final-cooling" channels~\cite{final-cooling} are also insensitive to the effect.
Modification of the Geant 4 variable-stepping agorithm to take this physics into account is under study. Until it is successfully implemented, G4beamline users are warned to include a study of stability with respect to variation of maxStep if they are using fields of $\sim$\,5\,T and above.

It is worth noting that the three simulations disagree at the $\pm$10\% level on the amount of scattering in 30\,cm of LH$_2$, with (for example) $x_{r.m.s.}=0.38$, 0.32, and 0.35\,mm (with negligible statistical uncertainty) for $B=30$\,T in the Gaussian, G4beamline, and ICOOL cases. This is not surprising, since the Bethe calculation~\cite{Bethe} and the PDG formula (Eq.~\ref{eq:scat}) are based on approximate (Thomas--Fermi) atomic potentials, whose use is unnecessary for hydrogen~\cite{Tollestrup}. G4beamline and ICOOL use more recent scattering models based on exact hydrogen wavefunctions~\cite{Fernow}. The residual 10\% disagreement between G4beamline and ICOOL was not investigated.

\ack 

We thank R. Fernow for providing the ICOOL example deck used in this study and for useful conversations. Work supported by U.S. D.O.E.

\end{document}